# Electrodynamic improvements to the theory of magnetostatic modes in ferrimagnetic spheres and their applications to saturation magnetization measurements

**Jerzy Krupka[1], Adam Pacewicz[2,*], Bartlomiej Salski[2], Paweł Kopyt[2], Jeremy Bourhill[3], Maxim Goryachev[3] and Michael Tobar[3]**

1) Warsaw University of Technology, Institute of Microelectronics and Optoelectronics, Koszykowa 75, 00-662 Warsaw, Poland
2) Warsaw University of Technology, Institute of Radioelectronics and Multimedia Technology, Nowowiejska 15/19, 00-665 Warsaw, Poland
3) ARC Centre of Excellence for Engineered Quantum Systems, University of Western Australia, 35 Stirling Highway, Crawley WA 6009, Australia

Abstract - Electrodynamic theory applied to the analysis of $TE_{n0p}$ mode resonances in ferromagnetic spheres placed either in metallic cavities or in the free space is compared with Walker-Fletcher's theory of so-called magnetostatic modes. The influence of the diameter of the sample, its permittivity and the permittivity of the surrounding media on the resonance frequencies of a few modes is analyzed. It is shown that the dominant resonances are essentially related either to negative values of the diagonal component of the permeability tensor or, for clockwise circularly polarized magnetic fields, to negative effective permeability. The electrodynamic theory is used to determine the saturation magnetization ($M_s$) from measured $TE_{n01}$ frequency differences. Measurements on different samples confirmed that $M_s$ can be determined using an electrodynamic approach with uncertainties of the order of 2% regardless of sample sizes, metal enclosures or static magnetic field values.

## 1. Introduction

As it is well known, at microwave frequencies several resonances can be excited in a ferromagnetic sphere biased with a static magnetic field. Such ferrimagnetic resonators find applications in tunable microwave filters, oscillators, and power limiters [1]. They usually operate on the dominant resonance mode, which is called the Kittel mode or the mode of uniform precession [2], which has recently been shown to be a magnetic plasmon mode [3]. Other higher-order modes are usually considered to be undesirable spurious responses that can be suppressed by decreasing the size of the sphere [2] or by an appropriate choice of the saturation magnetization ($M_s$). However, measurements of the resonance frequencies and the resonance linewidth of such higher-order modes as a function of the static magnetic field bias (or as a function of frequency at a fixed bias) allow to determine basic parameters of ferromagnetic materials, such as relaxation time of the precession mode under observation, magnetic anisotropy constants, the effective gyromagnetic ratio and $M_s$ [4]. The uncertainties of measurements of these parameters at microwave frequencies predominantly depend on the accuracy of the applied physical models. In this context, the purpose of this paper is two-staged. First, the relationship between magnetostatic and electrodynamic theories of the said higher-order modes will be elucidated. Second, it will be shown how the electrodynamic theory can be applied to microwave measurements of $M_s$ for improved accuracy as compared to methods that are based on the magnetostatic theory. A short but comprehensive summary of the state-of-the art is provided in Section 2, followed by extensive numerical results presented in Section 3, and subsequently by experimental results of $M_s$ measurements in Section 4.

## 2. Theory of ferromagnetic resonances – state of the art

The microwave magnetic properties of ferromagnetic materials magnetized above saturation along the $z$-direction of an appropriately chosen coordinate system can be quantitatively described with the Polder tensor $\bar{\bar{\mu}}$ derived from the Landau-Lifshitz equation [5], [6]:

$$\bar{\bar{\mu}} = \mu_0 \begin{bmatrix} \mu & j\kappa & 0 \\ -j\kappa & \mu & 0 \\ 0 & 0 & 1 \end{bmatrix} \quad (1)$$

For a lossless material, the tensor components take the form [7]:

$$\mu = 1 + \frac{H_{0r}}{H_{0r}^2 - w^2}, \quad (2)$$

$$\kappa = \frac{w}{H_{0r}^2 - w^2}, \quad (3)$$

where: $H_{0r} = H_0/M_S$, $w = f/f_m$, $f_m = \gamma M_S$, $H_0$ is the internal static magnetic field, $M_S$ is the saturation magnetization, $\gamma$ is the gyromagnetic ratio, and $f$ is the frequency. More details can be found in Appendix A.

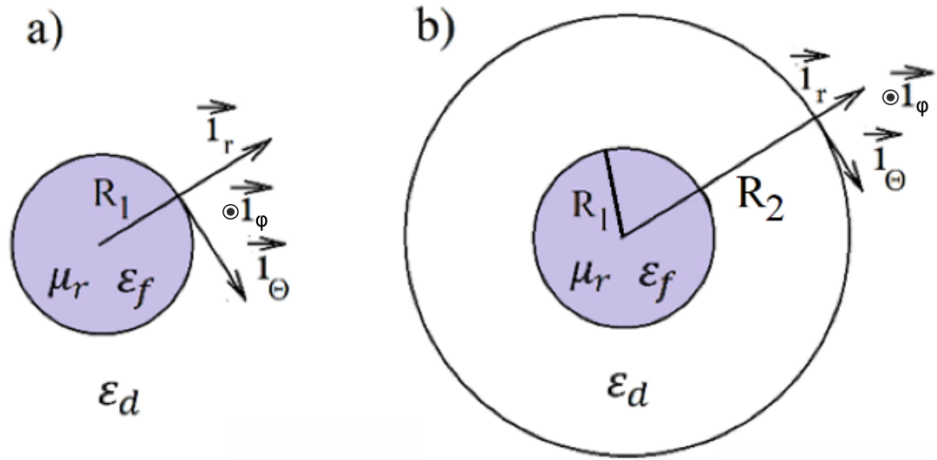

Fig.1. Electrodynamic models of an a) unshielded and b) shielded spherical ferrimagnetic resonator ($\vec{\iota}_r, \vec{\iota}_\phi, \vec{\iota}_\theta$ – versors of the spherical coordinate system).

Resonance frequencies of isolated ferromagnetic spheroids were analysed for the first time by Walker [2], who derived a transcendental equation (TDE) that allows to determine the relationship between the resonance frequency and the internal static magnetic field of an isolated ferromagnetic spheroid. The

TDE was derived under the assumption that the microwave magnetic fields in a gyromagnetic spheroid, magnetized along the axis of its revolution, satisfy the equations $\nabla \cdot \boldsymbol{B} = 0$ and $\nabla \times \boldsymbol{H} = 0$, which constitutes a magnetostatic approach. Later, P. C. Fletcher and R. O. Bell [8] derived a TDE for a gyromagnetic sphere as shown in Fig.1a, which is a special case of the TDE derived by Walker:

$$n + 1 + \xi_0 \frac{(P_n^m(\xi_0))'}{P_n^m(\xi_0)} \pm m\kappa = 0, \qquad (4)$$

where $\xi_0^2 = \frac{\mu}{\mu-1}$, $m$ ($n$) is the mode index related to the coordinate $\varphi$ ($\Theta$) shown in Fig.1, where $m \leq n$, and $P_n^m$ are the associated Legendre polynomials. For a given normalized internal magnetic bias, $H_{0r}$, Eq. (4) can be numerically solved with respect to the normalized frequency offset $\Delta w = w - H_{0r}$ (see open-source MATLAB code available in [9]) for any mode [$n$ $m$ $r$], where $r$ is the mode index related to the radial coordinate ($r = 0$ for the first solution).

Magnetostatic theory allows to accurately determine the resonance frequencies of fundamental and higher-order modes if the following conditions are satisfied:

1. The diameter of the ferromagnetic sphere, $d$, is much smaller than the free space wavelength at a given resonance frequency ($d \ll \lambda$). Otherwise, the magnetostatic model overestimates the resonance frequency. The influence of $d$ and $\varepsilon_f$ on the resonance frequency of the dominant mode ([1 1 0]), has been analysed using approximate methods [10], [11] and is given by [12]:

$$w - H_{0r} = \frac{1}{3} - \frac{4\pi^2}{90}(\epsilon_f + 5)\left(\frac{d}{\lambda}\right)^2 \qquad (5)$$

2. The sample is far enough from metal objects, such as cavity or strips, affecting the field distribution of the mode.
3. The ferromagnetic sphere is located in vacuum as the magnetostatic theory doesn't allow to consider other surrounding media.

However, it has been recently shown that the rigorous electrodynamic analysis of the mode of uniform precession lacks the aforementioned drawbacks [7]. In principle, all resonances of a magnetized ferromagnetic sphere whose microwave magnetic field is perpendicular to the direction of magnetization can be regarded as TE$_{n0p}$ modes occurring in a sphere of scalar relative permeability $\mu_r = \mu + \kappa$, enabling the formulation of a rigorous TDE provided that the coordinate system is appropriately chosen (see Appendix B). The mode index related to the $r$ ($\Theta$) coordinate shown in Fig.1 is $p$ ($n$). The azimuthal index, $m$, does not affect the resonance frequency and is therefore fixed at zero. The $r$-$\theta$ plane of the said coordinate system must rotate clockwise about the magnetization axis with the same radial frequency as the mode of interest rotates. In this nomenclature, the mode of uniform precession is denoted as the TE$_{101}$ mode [7]. In addition to much better accuracy of the electrodynamic TDE as compared to magnetostatic TDE, the former one also enables computations of the quality factor of the mode, which is related to the ferromagnetic linewidth [7].

Either of the discussed TDEs can be used for the determination of $M_s$ of ferromagnetic spheres. Dillon [13] and White [14] suggested to use for this purpose the frequency offset between different ferromagnetic modes. Numerous papers have been published demonstrating the practical measurement of $M_s$ based on Walker-Fletcher's magnetostatic theory, e.g. [15]–[17] and references therein. The general principle is that measurements of frequency spacing between two well identified modes $k$ and $l$ can be used to determine the saturation magnetization in the following way:

$$M_s = (f_k - f_l)/(w_k - w_l)/\gamma, \qquad (6)$$

where $f_i$ is the measured frequency and $w_i$ is the corresponding normalized frequency obtained with a chosen TDE. White [4] proposed to employ Eq. (5), which is approximate, to correct for propagation effects in the determination of the modes' frequency spacing. Still, the proposed corrections do not take the influence of the metal enclosure into account. In this paper, we propose to solve the problem rigorously by using the electrodynamic approach to determine the frequency spacing between different $TE_{n01}$ modes, as it lacks all the drawbacks of the magnetostatic method, enabling more accurate $M_s$ measurements even for higher bias fields.

## 3. Calculations

### 3.1. Resonance modes solved with the magnetostatic TDE

Extensive studies of resonances occurring in ferromagnetic spheres based on the magnetostatic TDE have been performed by P. Röschmann and H. Dötsch [18]. We will present the magnetostatic solution in a slightly different manner than it is known from the literature [2], [8], [18] by analyzing the permeability corresponding to selected resonances. It can be shown that the normalized frequency offset, $\Delta w = w - H_{0r}$, solved with Walker-Fletcher's TDE (see Eq. (4)), can be expressed as functions of the diagonal permeability, $\mu$, employed in the tensor given in Eq. (1) and effective permeability of the mode of uniform precession, $\mu_r = \mu + \kappa$, in the following way, respectively:

$$\Delta w^{-1} = (1 - \mu)^{-1}\left(1 + \frac{w}{H_{0r}}\right)^{-1}, \qquad (7)$$

$$\Delta w = (1 - \mu_r)^{-1}, \qquad (8)$$

where losses have been neglected. Fig. 2 shows the normalized frequency offsets, $\Delta w$, from the frequency where the pole of the Landau-Lifshitz-Gilbert permeability is present (Eq. (2)), for which the term FMR will be used. The shown $\Delta w$ offsets have been computed with the aid of the magnetostatic TDE. It should be emphasized that the mode of uniform precession occurring at $w = H_{0r} + \frac{1}{3}$ is commonly identified as the pole of Eq. (2), where material losses are at the maximum, which has recently been shown not to be the case [3].

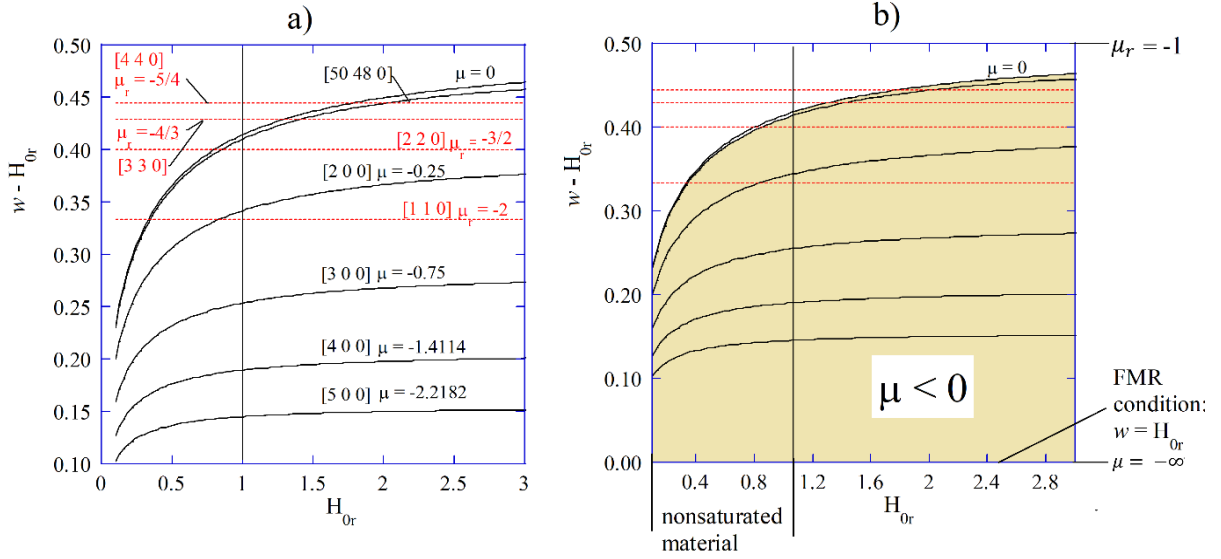

Fig.2. (a) Frequency offsets of selected modes from the FMR frequency computed using the magnetostatic TDE and (b) the corresponding range of negative $\mu$. Mode notation, [n m r], has been used after Fletcher [8].

As it can be seen in Fig. 2, for modes of the order [n, n, 0] the normalized frequency offset $\Delta w$ is constant and, therefore, w increases linearly with $H_{0r}$ for these modes. This statement is also true for [n, n-1, 0] modes [19], which have not been shown in Fig. 2. Both mode families are clockwise circularly polarized with $\mu_r = const < 0$ [19]. Different spectral properties are exhibited by modes of the order [n, 0, 0] for which $\Delta w$, and the corresponding w, increases with $H_{0r}$. In this case, the mode is uniquely defined with $\mu = const < 0$. Computations also show that $\mu < 0$ for the mode families [n, m, 0], $m \leq n - 2$, for a saturated material, as indicated with the shadowed region in Fig. 2b. For this family of modes, the resonance curves converge to the $\mu = 0$ limit as the m index increases, as shown in Fig. 2a for the [50, 48, 0] mode. It should also be noticed that the frequency at which the pole of the Landau-Lifshitz-Gilbert permeability is present ($\Delta w = w - H_{0r} = 0$), where no electromagnetic mode is present, has also been indicated in Fig. 2b.

In general, all the modes of any n and m indices, and r = 0, are located between the FMR level, $w = H_{0r}$ ($\mu = -\infty$), and $w = (H_{0r}(1 + H_{0r}))^{0.5}$ ($\mu = 0$) for a high enough $H_{0r}$. The effective permeability for the [n, n, 0] modes is given by the formula $\mu_r = -\frac{n+1}{n}$ and it converges to $\mu_r = -1$ for $n \to \infty$. Concluding, modes solved with the magnetostatic TDE are mostly related to the negative values of either the diagonal permeability $\mu$ or the effective permeability $\mu_r$.

## 3.2 Resonance modes solved with the electrodynamic TDE

In this section, we present $\Delta w(H_{0r})$ plots of the modes for the unshielded and shielded YIG sphere with d = 5 mm computed using the electrodynamic TDE and utilize electrodynamic mode notation ($TE_{n0p}$).

The diameter of the cavity was assumed to be $D_c = 10$ mm. Extensive measurements of resonances of a single-crystal spherical YIG sample of diameter $d = 5$ mm have been already performed [20]. The frequency offset of selected $TE_{n0p}$ modes obtained with the Walker-Fletcher model are plotted with a dashed line in Fig. 3a and Fig. 4a assuming $M_s = 140$ kA/m. The range of $H_{0r}$ in Fig. 3 roughly corresponds to the range of the external static magnetic induction varying from 0 to 1.8 T and resonant frequencies spanning up to 50 GHz. The $TE_{101}$ mode converges to $\mu_r = -2$ (i.e. $\Delta w = 1/3$) with decreasing $H_{0r}$, which agrees with the results obtained with Walker-Fletcher's TDE. However, $\Delta w$ decreases with increasing $H_{0r}$ (and with the corresponding frequency) quite rapidly. In case of the $TE_{101}$ mode, the effective permeability converges to $\mu_r = -\infty$ with increasing $H_{0r}$, which is the FMR condition. To the authors' knowledge, the electrodynamic TDE is the only known means of quantitatively predicting such behavior.

The frequency offset, $\Delta w$, increases in the presence of a metallic shield, especially for the lower order radial modes since the rate of evanescent decrease of the fields outside the sphere is smaller. It can be confirmed in Fig. 4b, which shows electric field distribution along the radius for a few $TE_{10p}$ modes obtained at $H_{0r}=2.5$ for the $TE_{101}$ mode and at $H_{0r}=5$ for the remaining $TE_{10p}$ modes. It is seen that the electric field of a $TE_{101}$ mode is also evanescent inside the sample ($\mu_r < 0$), while for other modes a standing wave is observed in that region ($\mu_r > 0$). For field distributions of $TE_{n01}$ modes, the reader is referred to our earlier work [3].

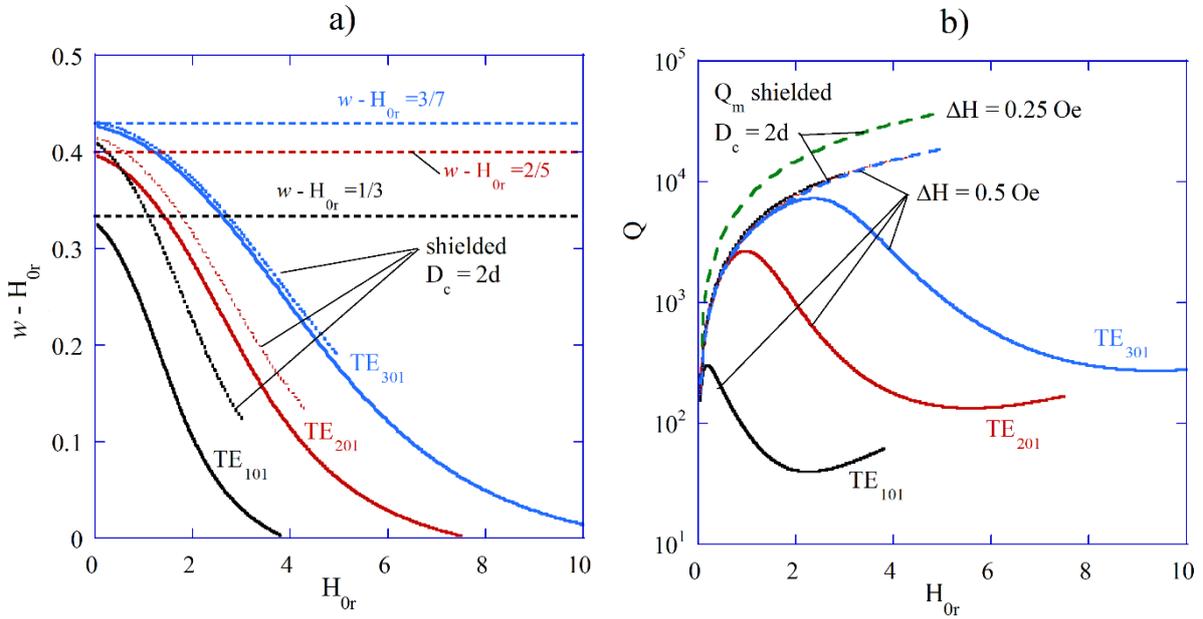

Fig.3. a) Frequency offset characteristics of selected modes solved with the magnetostatic (dashed line) and electrodynamic (solid and dotted lines) TDEs. Results have been obtained for a YIG sphere ($d = 5$ mm, $M_S = 140$ kA/m). Solid lines refer to the unshielded sphere, and dotted lines refer to the shielded sphere with $D_c = 10$ mm. b) Q-factors as a function of magnetic bias obtained with the electrodynamic TDE. For the shielded structure, it has been assumed that the shield is perfectly conducting.

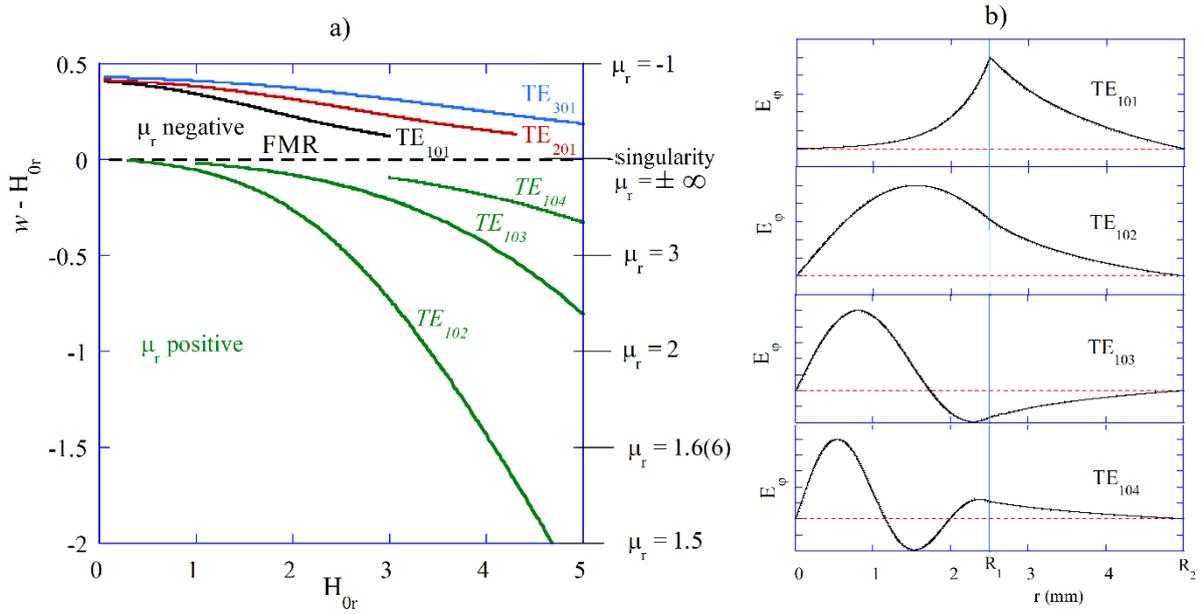

Fig.4. a) Frequency offset characteristics of selected modes solved with the electrodynamic TDE for a shielded YIG sphere ($d = 5$ mm, $D_c = 10$ mm, $M_S = 140$ kA/m). b) Plots of the electric field distribution for four $TE_{10p}$ modes. Plots of the electric field were obtained at $H_{0r}=2.5$ for the $TE_{101}$ mode and at $H_{0r}=5$ for the remaining three $TE_{10p}$ modes.

The Q-factor of a few TE modes is presented in Fig. 3b. For the unshielded sphere, the total Q-factor accounts for radiation and magnetic losses, while for the shielded structure, only magnetic losses are present assuming that the cavity is perfectly conducting. It can be noticed that the electrodynamic TDE allows to predict the frequency above which radiation losses exceed magnetic losses [21], and this frequency increases with the elevation mode order, $n$.

### 3.3 Electrodynamic analysis of factors influencing mode spacing

Frequency offsets obtained for $TE_{n01}$ modes, $n = 1 \ldots 3$, are shown in Fig. 5 for two diameters of the sample, while Fig. 6 presents the frequency spacing between the two modes $\Delta w_{21} = w_2 - w_1$. In qualitative accordance with Eq. (5), the resonance frequencies of $TE_{n01}$ modes decrease with increasing sample diameter $d$ and permittivity $\varepsilon_f$. As demonstrated in Fig. 5, increasing permittivity of the surrounding medium, $\varepsilon_d$, has a similar effect. The increase of either of the model parameters $H_{0r}$, $d$, $\varepsilon_f$, $\varepsilon_d$ lowers each consecutive $TE_{n01}$ mode frequency to a lesser degree, leading to the increase of $\Delta w_{n1}$ mode spacings. Similarly, as it is seen in Fig.6, the frequency spacing increases with the increase of the diameter of the metal enclosure, $D_c$. It was shown in [13] that the spacing between the modes is essentially independent of crystalline orientation. However, in general, since crystalline orientation influences the internal bias [7], it can influence the mode spacing, but it will be practically noticeable only for large anisotropy values.

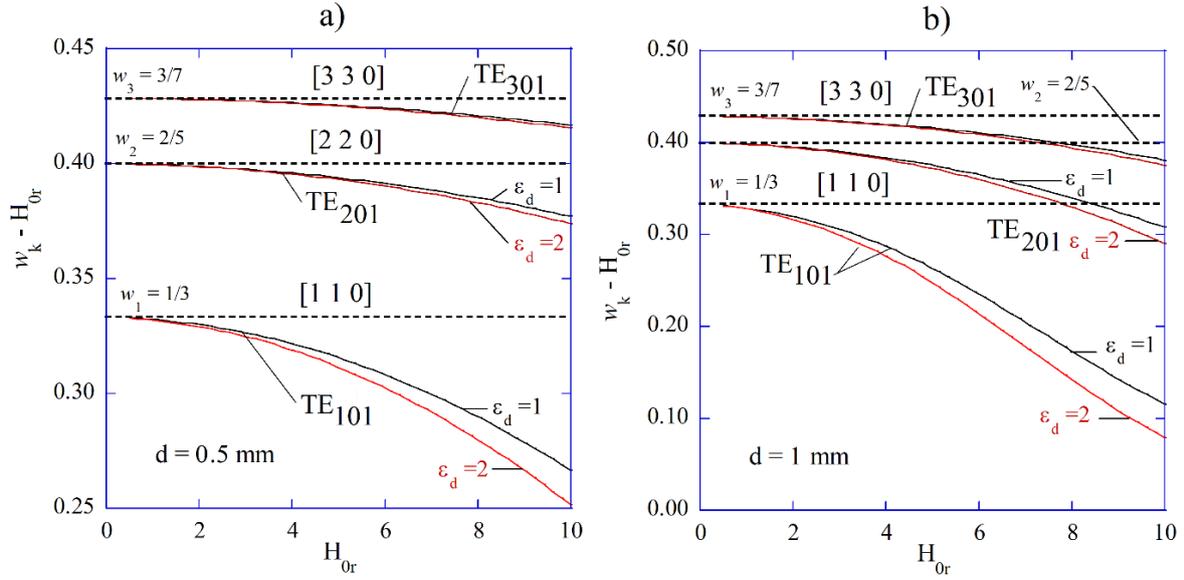

Fig.5. Differences between normalized frequencies and the normalized static magnetic field for $TE_{101}$, $TE_{201}$ and $TE_{301}$ modes as a function of the normalized static magnetic field for two different permittivity values of medium surrounding ferromagnetic samples ($\varepsilon_d$) for samples having diameter equal to a) $d$ =0.5 mm, and b) $d$ =1.0 mm.

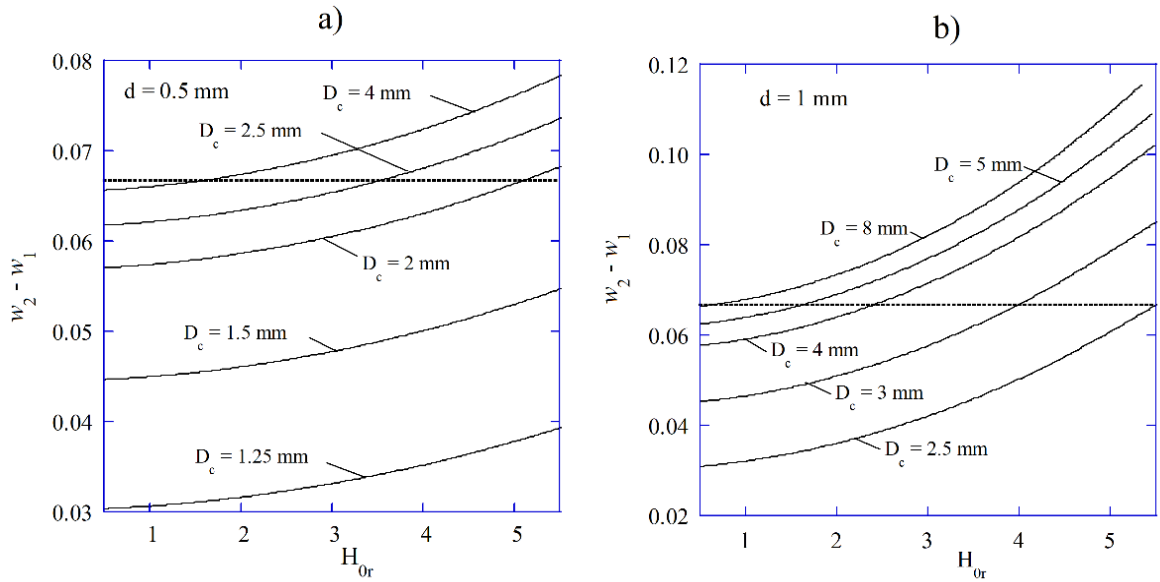

Fig.6. Differences between normalized frequencies of $TE_{201}$ ($w_2$) and $TE_{101}$ ($w_1$) modes as functions of normalized static magnetic field for various diameters ($D_c$) of metal enclosures. $M_s$ = 140 kA/m, $\varepsilon_f$ = 16, $\varepsilon_d$ = 1 were assumed. a) sample diameter $d$ =0.5 mm, b) sample diameter $d$ =1.0 mm.

## 4. Experiments

Experiments have been performed in the setup shown in Fig.7. The single-crystal YIG sphere ($d$ = 0.5 mm) in a PTFE tube is inserted between the orthogonally oriented coupling loops connected to a vector network analyzer (VNA), which is employed to measure the scattering matrix coefficient ($S_{21}$). The loops surrounding the sample are soldered to the metal enclosure made of copper ($D_c$ = 2.5 mm). An electromagnet was used as a source of static magnetic bias. Fig. 8 shows an exemplary spectrum of ferromagnetic modes with $TE_{101}$, $TE_{201}$ and $TE_{301}$ modes annotated. Such spectra were recorded for several static magnetic bias values.

As is postulated in this paper, considering variations of the frequency spacing between the $TE_{n0p}$ modes using the electrodynamic TDE, it is possible to improve the accuracy of the determination of $M_s$. The procedure to obtain $M_s$ from the set of measured resonant frequencies $f_k, f_l$ obtained for different external magnetic bias values $H^k_{ext}$ is as follows:

1. As a starting point, take an initial value of $M_s$ based on knowledge of the material or determine $M_s$ using Eq. (6) from $f_k, f_l$ measured at low static bias assuming magnetostatic values of $w_k$ and $w_l$.
2. Compute $w_k$ and $w_l$ for each $H^k_{ext}$ using the electrodynamic TDE.
3. Compute $M_s$ for each $H^k_{ext}$ using Eq. (6).
4. Repeat the procedure starting from the average value of $M_s(H_{ext})$ until convergence.

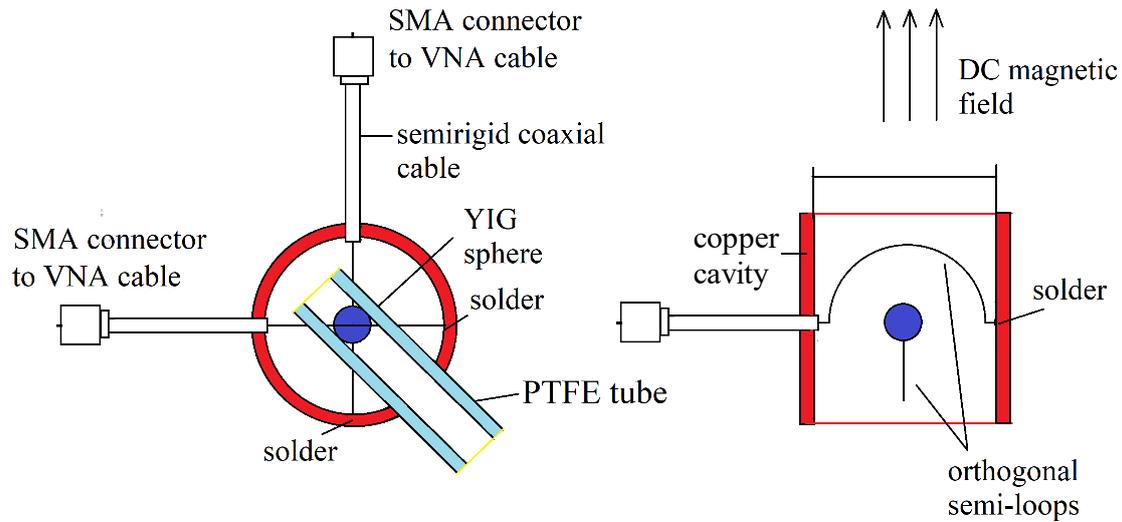

Fig.7. Schematic of experimental setups – top view (left) and side view (right).

Table I shows measured frequency spacing for the spectrum shown in Fig. 8 and the corresponding $M_s$ determined with the aid of both TDEs. It is seen that the normalized frequency differences obtained employing the electrodynamic model are smaller than their magnetostatic counterparts, which confirms theoretical results shown in Fig. 6a. Values of $M_s$ obtained with the aid of the electrodynamic and magnetostatic TDEs using Eq. (6) well agree with the nominal value provided by the manufacturer ($M_S = 139.2$ kA/m) if one considers the spacing between the modes $TE_{301}$ and $TE_{201}$. However, the magnetostatic model gives rise to about 7 % of error in $M_S$ if $TE_{201}$ and $TE_{101}$ modes are analyzed due to the erroneous prediction of $\Delta w_{21}$ at the given magnetic bias. The use of the latter pair of modes has the practical advantage that it is easier to couple to the $TE_{201}$ than to the $TE_{301}$ mode.

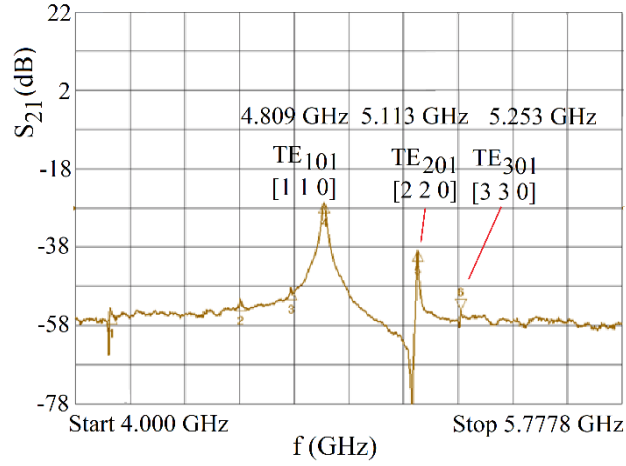

Fig.8. Spectrum of $|S_{21}|$ in the setup shown in Fig. 7 for a YIG sphere ($d$ = 0.5 mm, $M_S$ =139.2 kA/m) at $H_{0r} \approx$ 0.65.

Table I. Saturation magnetization determined for a YIG sample ($d$ = 0.5 mm) in metal enclosure ($D_c$ = 2.5 mm) from measured resonance frequencies and the corresponding computed frequency offsets. The static magnetic bias was such that $H_{0r} \approx 0.65$.

| $k$ | $f$ (MHz) | $\Delta w_{k,k-1}$ magnetoststic | $\Delta w_{k,k-1}$ electrodynamic | Fletcher's mode labelling | Electrodynamic $TE_{n0p}$ mode designation | $M_s$ (kA/m) electrodynamic | $M_s$ (kA/m) magnetostatic |
|---|---|---|---|---|---|---|---|
| 1 | 4809 | | | 1 1 0 | 1 0 1 | | |
| 2 | 5113 | 0.066667 | 0.062096 | 2 2 0 | 2 0 1 | 139.04 | 129.51 |
| 3 | 5253 | 0.028571 | 0.028568 | 3 3 0 | 3 0 1 | 139.18 | 139.24 |

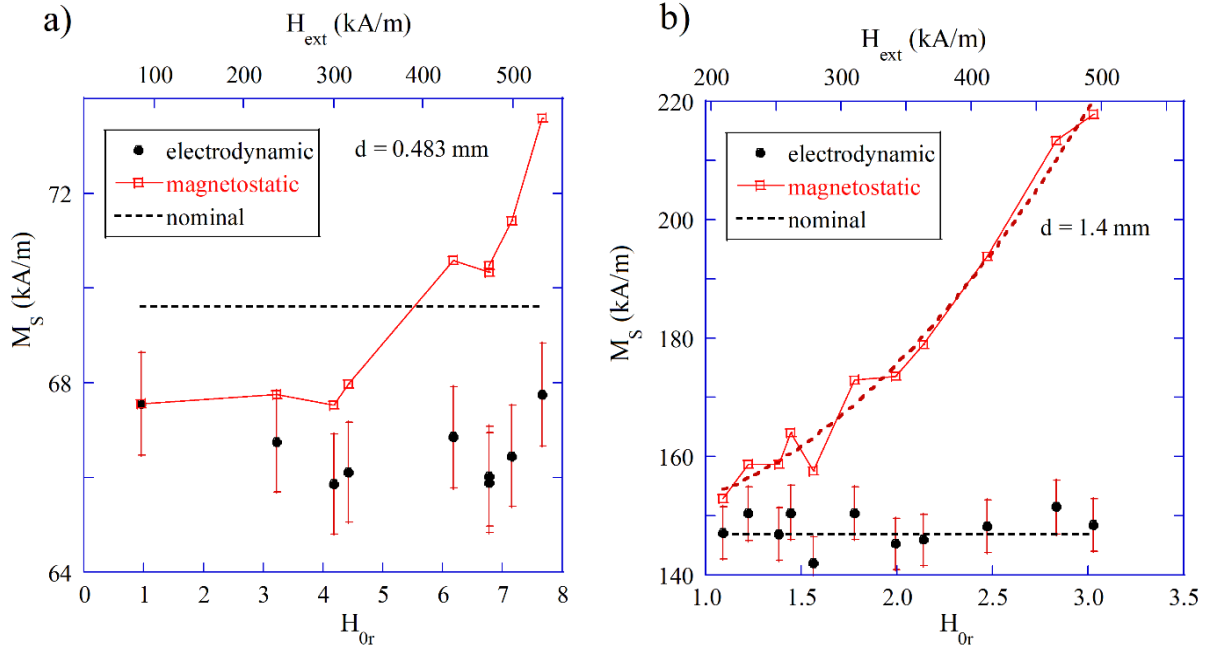

Fig.9. Saturation magnetization, $M_s$, determined from the frequency offset between $TE_{101}$ and $TE_{201}$ modes as a function of magnetic bias obtained for a) a single-crystal doped YIG ($d = 0.483$ mm, $\Delta H \approx 1$ Oe at 10 GHz), and b) polycrystalline calcium vanadium ferrite ($d = 1.4$ mm, $\Delta H \approx 4$ Oe at 10 GHz). The larger relative errors in Fig. 9b as compared to Fig. 9a arise because multiple modes are excited in the larger sphere and they influence each other's resonance frequency due to moderate mode $Q$-factors.

Fig.9 shows the saturation magnetization determined from the frequency offsets between $TE_{101}$ and $TE_{201}$ modes measured in a broad range of magnetic bias for two additional ferromagnetic samples. The first sample (Fig. 9a) is made of single-crystal Ga-doped YIG having $d = 0.483$ mm and $\Delta H \approx 1$ Oe at 10 GHz. The second sample (Fig. 9b) is made of polycrystalline calcium vanadium ferrite having $d = 1.4$ mm and $\Delta H \approx 4$ Oe at 10 GHz. Samples were measured in the experimental setup shown in Fig. 7. For the smaller sample, $M_s$ obtained with the proposed electrodynamic approach amounted to ca. 66.6 kA/m within a 2% relative error, which is lower than the nominal value of 69.6 kA/m declared by Vendor #1. Calculating $M_s$ using the magnetostatic approach leads to an overestimation of $M_s$ by up to ca. 10% compared to the electrodynamic approach. For the larger calcium vanadium sample having $d = 1.4$ mm, computed values of $\Delta w_{21}$ and measured frequency offsets between the corresponding modes increase with $H_{0r}$ by as much as 50% in the measurement range. As a result, the saturation magnetization determined with Eq. (6) with the aid of the electrodynamic TDE remains constant within an experimental error of a few per cent (see Fig. 9b) and is in close agreement with the nominal value of 147.2 kA/m declared by Vendor #2. On the contrary, the saturation magnetization determined from the magnetostatic TDE, where $\Delta w_{21} = \frac{2}{5} - \frac{1}{3} = const$, increases up to about 50% above the nominal value at $H_{0r} = 3$.

## 5. Discussion

Analysis of the methods applicable to the measurement of $M_s$ using Eq. (6) available in the literature [15] reveals that experimental errors can reach even +58% if the magnetostatic approach is used. The presented experiments, thus, confirm the superiority of the electrodynamic approach as the error in the measurement of $M_s$ predominantly depends on the accuracy of the theoretical assessment of $w_k$ and $w_l$. Contrary to experiments, these coefficients obtained with the magnetostatic model are constant with $H_{0r}$ and other parameters. It should be mentioned that the scattering theory (see Eq. (5)) allows to approximate the resonance frequency changes of the mode of uniform precession as function of sample size and its permittivity in an infinitely large metal enclosure and up to $d/\lambda \approx 0.1$ [12]. However, this theory is not strictly applicable for the higher order modes. Therefore, the electrodynamic model is a natural choice for accurate computations of $w_k - w_{k-1}$ and, consequently, accurate determination of $M_s$. The use of the electrodynamic TDE instead of Walker-Fletcher's TDE allows to remove many practical sources of error present in the determination of $M_s$ with Eq. (6) and extend the measurement range to higher static bias fields.

## 6. Conclusions

Results of our computations of $TE_{n0p}$ modes in shielded and unshielded ferromagnetic spheres employing a rigorous electrodynamic approach confirm that these modes can be treated as ordinary electromagnetic resonances in a gyrotropic, dispersive and lossy magnetic medium. The magnetostatic model of these resonances allows to analyze only some of their properties, most notably resonance frequencies for sufficiently small samples. Most of the dominant resonances that exist in small ferromagnetic spheres below the frequency of FMR are related either to the negative effective permittivity value ($\mu_r$) for circularly polarized [$n,n,0$] (also known as $TE_{n01}$ using electrodynamic notation) and [$n,n-1,0$] modes or to the negative value of the diagonal component $\mu$ of permeability tensor. The presented electrodynamic theory allows for accurate measurements of the basic parameters of magnetic materials including, but not limited to, the saturation magnetization.

**Acknowledgement**

This work was partially supported by the TEAM-TECH 2016-1/3 Project entitled "High-precision techniques of millimeter and sub-THz band characterization of materials for microelectronics" operated within the Foundation for Polish Science TEAM TECH Program co-financed by the European Regional Development Fund, Operational Program Smart Growth 2014-2020.

This work was partially supported by "Project PROM - International Scholarship Exchange for PhD Students and Academic Staff", financed from the European Social Fund under the Operational Program Knowledge Education Development, non-competitive project entitled International Scholarship Exchange for PhD Students and Academic Staff, contract number: POWR.03.03.00-00-PN13/18.

This work was partially supported by "The ARC Centre of Excellence for Engineered Quantum Systems", project grant number CE170100009.

This work was partially supported by the National Science Centre, project registration number 2018/31/B/ST7/04006.

**Appendix A**

**Permeability tensor**

The ferromagnetic resonance (FMR) phenomenon (FMR) can be quantitatively described with a permeability tensor $\bar{\bar{\mu}}$ derived from the Landau-Lifshitz equation [6]. In a presence of uniform static magnetic field magnetizing ferrite material along $z$-axis of Cartesian or cylindrical coordinate system the permeability tensor takes the form (1a). When the static magnetic field is sufficiently strong to saturate gyromagnetic medium, then $\mu_{||} = 1$ and permeability tensor is known as the Polder tensor [5].

$$\bar{\bar{\mu}} = \mu_0 \begin{bmatrix} \mu & j\kappa & 0 \\ -j\kappa & \mu & 0 \\ 0 & 0 & \mu_{||} \end{bmatrix} \quad (1a)$$

The diagonal and the off-diagonal relative components of the Polder tensor take the following form [6]:

$$\mu = 1 + \frac{H_{0r} + j\,\alpha\,\widehat{w}}{H_{0r}^2 - \widehat{w}^2 + 2\,j\,\alpha\,H_{0r}\,\widehat{w}} \quad (2a)$$

$$\kappa = \frac{\widehat{w}}{H_{0r}^2 - \widehat{w}^2 + 2\,j\,\alpha\,H_{0r}\,\widehat{w}} \quad (3a)$$

where: $H_{0r} = H_0/M_S$, $\widehat{w} = \hat{f}/f_m$, $f_m = \gamma M_S$, $H_0$ is the static magnetic field inside the sample (the internal static magnetic field), $M_S$ is the saturation magnetization, $\alpha$ is a Gilbert damping factor, and $\hat{f}$ is the complex frequency.

The natural frequency of ferromagnetic resonance $f_m$ is defined as $f_m = \gamma_{eff} M_S$ where $M_S$ is the saturation magnetization of ferrite material, and $\gamma_{eff}$ is the effective gyromagnetic ratio. For free electron gyromagnetic ratio is known with high precision from cyclotron measurements $\gamma_e = \frac{|e|}{2m_e} g_e \approx 35.21719$ MHz/(kA/m), and $g$-factor $g_e = 2.002319$, while for real ferrite materials $g$-factor values are determined from experiments. $g$-factor values for typical microwave ferrites are close to $g_e$ so it is often assumed that $\gamma_{eff} = \gamma_e = \gamma$.

The imaginary part of frequency describes time dependence of electromagnetic fields (transient solutions) for a microwave resonator containing lossy medium. Instead to the Gilbert damping factor the relaxation time, $\tau = 1/(\alpha \gamma H_0)$, and the ferromagnetic resonance linewidth, $\Delta H = 2\alpha H_0 = 2/(\gamma \tau)$, are often used to describe the losses in ferromagnetic material.

For circularly polarized electromagnetic fields permeability of ferromagnetic material can be expressed as (4).

$$\mu_{r,l} = \mu \pm \kappa \quad (4a)$$

A positive (negative) sign in Equation (4a) corresponds to the clockwise (counter-clockwise) polarization of the magnetic field.

For lossless medium formulas (2a), (3a) and (4a) reduce to

$$\mu = 1 + \frac{H_{0r}}{H_{0r}^2 - w^2} \quad (2b)$$

$$\kappa = \frac{w}{H_{0r}^2 - w^2} \quad (3b)$$

$$\mu_r = 1 + \frac{1}{H_{0r} - w} \quad (4b)$$

Ferromagnetic resonance frequency $f_{FMR}$ is defined as the frequency for which the denominators in expressions (2a) and (3a) and (4a) vanish, which takes place when $f_{FMR} = \gamma H_0$ or $w = H_{0r}$.

## Appendix B

**Transcendental equations of electrodynamics**

For the TE$_{n0p}$ modes of free oscillations of an isotropic sphere having permeability $\mu_r$, relative complex permittivity $\varepsilon_f$, and immersed in dielectric medium having permittivity $\varepsilon_d$, as in Fig.1a, the TDE can be written as follows [3]:

$$\left\{nJ_{n+\frac{1}{2}}(kR_1) - kJ_{n-\frac{1}{2}}(kR_1)\right\}H^{(2)}_{n+\frac{1}{2}}(k_0R_1) - \mu_r\left\{nH^{(2)}_{n+\frac{1}{2}}(k_0R_1) - k_0H^{(2)}_{n-\frac{1}{2}}(k_0R_1)\right\}J_{n+\frac{1}{2}}(kR_1) = 0 \quad (6a)$$

$n, p$ are elevation (related to $\Theta$ in Fig.1) and radial mode indices, respectively, $c$ is the speed of the EM wave in a vacuum, $J$ ($H$) are Bessel (Hankel) functions. The radial mode index $p$ denotes consecutive roots of (6a) while the azimuthal mode index $m$ has no impact on the resonance frequency, so it is assumed to be $m = 0$.

For the TE$_{n0p}$ modes of free oscillations of an isotropic sphere having permeability $\mu_r$, relative complex permittivity $\varepsilon_f$, and immersed in dielectric medium having permittivity $\varepsilon_d$, as in Fig.1b, the TDE can be written as follows [8]:

$\det(W(\omega)) = 0 \qquad (7a)$

where: $W = \begin{bmatrix} w_{11} & w_{12} & w_{13} \\ w_{21} & w_{22} & w_{23} \\ 0 & w_{32} & w_{33} \end{bmatrix}$

$w_{11} = \mu_r(k_1R_1)^{\frac{1}{2}}J_{n+\frac{1}{2}}(k_1R_1)$

$w_{12} = -(k_2R_1)^{\frac{1}{2}}J_{n+\frac{1}{2}}(k_2R_1)$

$w_{13} = -(k_2R_1)^{\frac{1}{2}}Y_{n+\frac{1}{2}}(k_2R_1)$

$w_{21} = (k_1R_1)^{\frac{3}{2}}\left[J_{n-\frac{1}{2}}(k_1R_1) - \frac{n}{k_1R_1}J_{n+\frac{1}{2}}(k_1R_1)\right]$

$w_{22} = -(k_2R_1)^{\frac{3}{2}}\left[J_{n-\frac{1}{2}}(k_2R_1) - \frac{n}{k_2R_1}J_{n+\frac{1}{2}}(k_2R_1)\right]$

$w_{23} = -(k_2R_1)^{\frac{3}{2}}\left[Y_{n-\frac{1}{2}}(k_2R_1) - \frac{n}{k_2R_1}Y_{n+\frac{1}{2}}(k_2R_1)\right]$

$$w_{32} = (k_2R_2)^{\frac{1}{2}}J_{n+\frac{1}{2}}(k_2R_2)$$

$$w_{33} = (k_2R_2)^{\frac{1}{2}}Y_{n+\frac{1}{2}}(k_2R_2)$$

where:

$k_1 = \omega(\varepsilon_f\mu_r)^{0.5}/c$, $k_2 = \omega(\varepsilon_d)^{0.5}/c$

$Y$ – Bessel functions of the second kind.

When equations (6a) or (7a) are solved with respect to the complex angular frequency $\widehat{\omega}$, the $Q$-factors can be determined as $Q = \omega'/2\omega''$.